\documentclass[pre,twocolumn,preprintnumbers,amsmath,amssymb,nofootinbib,floatfix]{revtex4}

\usepackage{graphicx,bm, tikz, mathrsfs, braket}
\usepackage[breaklinks]{hyperref}

\makeatletter
\def\graphicscale{\twocolumn@sw{0.3}{0.4}}
\def\graphicthreescale{\twocolumn@sw{0.3}{0.4}}

\begin{document}

\title{Thermal-bath effects in quantum quenches within quantum critical
  regimes}

\author{Francesco Tarantelli}
\affiliation{Dipartimento di Fisica dell'Universit\`a di Pisa
        and INFN, Largo Pontecorvo 3, I-56127 Pisa, Italy}

\author{Ettore Vicari} 
\affiliation{Dipartimento di Fisica dell'Universit\`a di Pisa,
  Largo Pontecorvo 3, I-56127 Pisa, Italy}

\date{\today}

\begin{abstract}
  We address the out-of-equilibrium dynamics arising from
  quantum-quench (QQ) protocols (instantaneous changes of the
  Hamiltonian parameters) in many-body systems within their quantum
  critical regime and in contact with (homogeneously coupled) thermal
  baths. We consider two classes of QQ protocols.  In one of them the
  thermal bath is used to prepare the initial finite-temperature Gibbs
  state; then, after quenching, the thermal bath is removed and the
  dynamics of the system is unitary.  We also address a more complex
  QQ protocol where the thermal bath is not removed after quenching,
  thus the quantum evolution is also driven by the interaction with
  the bath, which may be described by appropriate master equations for
  the density matrix of the system, where a further relevant time
  scale, or inverse decay rate, characterizes the system-bath
  coupling.  Under these QQ protocols, the critical system develops
  out-of-equilibrium scaling behaviors, which extend those for
  isolated critical systems, by introducing further scaling variables
  proportional to the temperature of the thermal bath and the decay
  rate of the system-bath interactions.  These out-of-equilibrium
  scaling behaviors are checked by analyzing QQ protocols within
  fermionic Kitaev wires, or equivalently quantum Ising chains,
  supplemented with a particular modelization of thermal bath that
  guarantees the asymptotic thermalization within the Lindblad master
  equation for the dynamics of open systems.
\end{abstract}

\maketitle


\section{Introduction}
\label{intro}

Thanks to the recent experimental progress in the realization and
control of the dynamics of quantum many-body systems, see
e.g. Refs.~\cite{Bloch-08, GAN-14}, the out-of-equilibrium quantum
dynamics of many-body systems has become an important theoretical
issue. In particular, out-of-equilibrium phenomena have been addressed
within the critical regimes of many-body systems at continuous quantum
transitions (CQTs)~\cite{SGCS-97,Sachdev-book,RV-21}, where collective
behaviors give rise to zero-temperature singularities in the
equilibrium low-energy properties of the system, and the universal
critical behaviors are determined by a limited number of relevant
features, such as the global symmetry, the symmetry-breaking pattern,
dimensionality, etc.. Within critical regimes and in the appropriate
thermodynamic or finite-size scaling (FSS) limits, one can achieve a
complete characterization of the complex dynamics of many-body systems
by controlling a limited number of renormalization-group (RG)
perturbations.  The universal scaling behaviors at CQTs extend beyond
the equilibrium conditions~\cite{RV-21}.  Indeed dynamic protocols
entailing out-of-equilibrium evolutions develop scaling behaviors as
well, in the appropriate limits, related to the universality class of
the CQT.  For example, out-of-equilibrium scaling behaviors emerge
when analyzing the quantum evolutions arising from a quantum quench
(QQ), see e.g. Refs.~\cite{BDD-15, RMD-17, HPD-18, PRV-18, TIGGG-19,
  RV-21, PV-23}, or from slow changes of the Hamiltonian parameters
across the transition point, such as the protocols associated with the
so-called quantum Kibble-Zurek problem, see
e.g. Refs.~\cite{Kibble-76,Kibble-80, Zurek-85,Zurek-96,ZDZ-05,
  PG-08,Dziarmaga-10,Dutta-etal-book, PSSV-11,CEGS-12,
  RDZ-19,RV-21,DV-23}.

These out-of-equilibrium issues have been mostly addressed within
isolated many-body systems, unitarily driven by their Hamiltonian and
the Schr\"odinger equation. In this paper we extend such studies to
investigate how the interaction with a thermal bath, coupled
homogeneously to the system, affects the out-of-equilibrium dynamics
of many-body systems within the critical regime of a zero-temperature
quantum transition, such as that arising from a QQ or a slow crossing
of the quantum critical regime.

The role of the temperature within the equilibrium critical behavior
at a CQT is generally associated with one of the relevant RG
perturbations at the stable fixed point of the RG flow controlling the
quantum criticality~\cite{SGCS-97,Sachdev-book,CPV-14,RV-21}.
Therefore, the quantum scaling behavior can be only observed in the
zero-temperature limit. More precisely, the quantum scaling limit
requires that the zero-temperature critical point is approached
keeping the ratio $T/\Delta$ fixed, where $\Delta$ is the gap at the
quantum critical point, which is generally power-law suppressed.  For
example, in the FSS limit the gap is suppressed as $\Delta\sim L^{-z}$
at the critical point, where $L$ is the size of the system and $z>0$
is the universal dynamic exponent associated with universality class
of the CQT. Within the equilibrium critical regime the temperature
enters the asymptotic FSS laws through a further dependence of the
scaling functions on the scaling variable $\Xi \equiv T L^z \sim
T/\Delta$.

The role of the temperature becomes less definite when we consider
out-of-equilibrium behaviors, because the temperature of the system is
an equilibrium concept. However, one may consider the effects of
thermal baths in contact with the system during its out-of-equilibrium
dynamics. The main feature of a thermal bath is that it eventually
drives the system toward thermalization at its temperature $T$, in the
large-time limit of the evolution of the system in contact with the
thermal bath.  The thermalization process must somehow introduce a
further time scale $\tau$ in the problem, characterizing the approach
of the system to the thermal state when it is put in contact with the
thermal bath. Such time scale is expected to play an inportant role in
the out-of-equilibrium dynamics of the system in contact with the
thermal bath.  In this paper we investigate these issues within the
simplest dynamic protocols giving rise to out-of-equilibrium
behaviors, i.e. those entailing instantaneous QQs of the Hamiltonian
parameters starting from equilibrium thermal conditions.

A quench protocol is generally performed by suddenly varying a
parameter within a family of Hamiltonians, such as
\begin{equation}
  \hat{H}(w) = \hat{H}_c + w \hat{H}_p,
  \label{qudef}
\end{equation}
where $\hat{H}_c$ and $\hat{H}_p$ are independent of the parameter
$w$, and $[\hat{H}_c , \hat{H}_p] \neq 0$.  In a standard QQ protocol
for closed systems, one usually starts from the ground state $|\Phi_0,
w_i\rangle$ of the Hamiltonian $\hat H(w_i)$ associated with an
initial value $w_i$ of the parameter $w$, with corresponding density
matrix $\rho_i = |\Phi_0, w_i\rangle \langle \Phi_0, w_i|$. At a given
time, $t=0$ say, the Hamiltonian parameter is suddenly changed from
$w_i$ to $w \neq w_i$, and the subsequent quantum evolution is
supposed to be unitarily driven by the Hamiltonian $\hat H(w)$, that
is $|\Psi(t)\rangle = e^{-i \hat H(w) t} |\Phi_0, w_i\rangle$
(hereafter we set $\hbar = 1$).  Several interesting issues have been
investigated within QQ dynamic protocols.  They include the long-time
relaxation and the consequent spreading of quantum correlations and
entanglement, the statistics of the work, localization effects due to
the mutual interplay of interactions and disorder, dynamical phase
transitions, the dynamic scaling close to quantum transitions, effects
of dissipation or of measurements due to interactions with an
environment (see, e.g., Refs.~\cite{Niemeijer-67, BMD-70, SPS-04,
  DMCF-06, SHLVS-06, RDYO-07, RDO-08, ZPP-08, PZ-09, IR-11, RI-11,
  GS-12, CEF-12a, CEF-12b, BRI-12, CE-13, HPK-13, FCEC-14, FHS-14,
  Cardy-14, NH-15, CTGM-16, CC-16, BD-16, IMPZ-16, LGS-16, VM-16,
  NRVH-17, Heyl-18, PRV-18, NRV-19-wo, NRV-19-cd, STT-20, RV-20,
  CLSV-20, PRV-20, RCFG-21, RV-21, GMEHB-02, KWW-06, HLFSS-07,
  Trotzky-etal-12, Cheneau-etal-12, Gring-etal-12, Schreiber-etal-15,
  Braun-etal-15, PCV-15, Kaufman-etal-16, Smith-etal-16, BLSKB-17,
  Zhang-etal-17, TNDTT-17, MCMJA-18, JJLM-19, Kohlert-etal-19,
  Maier-etal-19}).

To focus on the out-of-equilibrium dynamics close to a quantum
transition, we assume that the Hamiltonian $\hat{H}_c$ in
Eq.~(\ref{qudef}) is critical, thus $w=w_c=0$ represents a quantum
critical point.  We recall that the critical behavior around the CQT
point $w_c=0$ is characterized by a diverging length scale $\xi\sim
|w|^{-\nu}$ of the quantum critical modes, and the power-law
suppression $\Delta \sim \xi^{-z}$ of the gap. The out-of-equilibrium
dynamics at CQTs develops scaling behaviors controlled by the
universality class of the quantum transition, for example when the
Hamiltonian parameters are slowly varied across the critical
regime~\cite{CEGS-12, RV-21, DV-23}, and in the case of {\em soft} QQ
protocols when both the initial and final values of the quenching
parameters are such to maintain the system within the critical
regime~\cite{PRV-18, PRV-20, RV-21}. In particular, soft QQs require
that the energy scale of the QQ [i.e. the difference of the energy
  $\langle \Psi(t)|\hat{H}(w)|\Psi(t)\rangle$ of the evolving state
  $|\Psi(t)\rangle$ for $t>0$ and the ground state of $\hat{H}(w)$] is
sufficiently small, i.e. comparable with the energy gap $\Delta \sim
L^{-z}$ of the spectrum at the transition point in finite-size
systems.

To study the effects of a thermal bath in the out-of-equilibrium
behavior arising from a QQ within the critical regime, we consider two
protocols where the thermal baths are involved in different ways:

  (i) Within the first protocol the thermal bath is used to prepare
the system in a finite-temperature Gibbs state, described by the
thermal density matrix (hereafter we set the Boltzmann constant
$k_B=1$)
\begin{equation}
\rho_t(w_i,T) = \sum_n e^{-E_n(w_i)/T}|\Phi_n, w_i\rangle \langle
\Phi_n, w_i|,
\label{rhoi}
\end{equation}
where $|\Phi_n, w_i\rangle$ are the eigenstates of $\hat{H}(w_i)$.
Then the quantum evolution after the quench of the Hamiltonian
parameters at $t=0$ is unitary and driven by the Hamiltonian
$\hat{H}(w)$ only, i.e., the thermal bath is removed during the
quantum evolution for $t>0$. Therefore, the evolution of the density
matrix is driven by the equation
\begin{equation}
  \partial_t \rho(t) = - i [\hat{H}(w),\rho(t) ],\qquad \rho(t=0) =
  \rho_t(w_i,T).
  \label{firstprot}
\end{equation}

(ii) In the second protocol the starting point is the same, i.e.  the
Gibbs state (\ref{rhoi}), but the thermal bath is not removed after
quenching. Therefore, the out-of-equilibrium quantum evolution for
$t>0$ is not unitary anymore, but it is also driven by the interaction
with the thermal bath. Under some conditions, discussed in
Refs.~\cite{Lindblad-76, GKS-76,BP-book, RH-book, SBD-16,RV-21,DR-21},
the nonunitary evolution arising from the thermal baths can be
described by a Lindbald master equation governing the time evolution
of the density matrix of the system, which can be written as
  \begin{eqnarray}
    \partial_t \rho = \mathcal{L}[\rho]\equiv
    -i\bigr[\hat{H}(w),{\rho}\bigr]+ \gamma \, \mathbb{D}_T[\rho],
    \label{Lindblad}
  \end{eqnarray}
  where $\mathcal{L}$ is a Liouvillian superoperator, and
  $\mathbb{D}_T$ is a dissipative driving whose strength is controlled
  by the homogeneous coupling $\gamma$, playing the role of the decay
  rate (inverse time scale) associated with the interactions between
  the system and the bath.  The operator $\mathbb{D}_T$ is assumed to
  be such that the Lindbald master equation (\ref{Lindblad}) drives
  the system toward an equilibrium Gibbs state at temperature $T$ in
  the large-time limit.
  
  We argue that, for both types of protocols and sufficiently small
  temperatures of the thermal baths, the out-of-equilibrium time
  evolution within the critical regime develop a nontrivial
  out-of-equilibrium FSS (OFSS) limit, with peculiar scaling
  behaviors, similar to those arising for closed systems. The effects
  of the thermal baths can be taken into account by appropriate
  extensions of the out-of-equilibrium zero-temperature scaling laws
  describing soft quantum QQs within the critical regime of isolated
  systems, already put forward by earlier works~\cite{PRV-18,RV-21}.
  As a theoretical laboratory to check our extended OFSS laws, we
  consider the quantum Ising chain~\cite{Sachdev-book}, or the
  equivalent fermionic Kitaev wire~\cite{Kitaev-01}, supplemented with
  a particular modelization of the thermal bath that guarantees the
  asymptotic thermalization within the Lindblad formulation of the
  dynamics of open systems with quadratic
  Hamiltonians~\cite{DR-21,PCR-22}, such as the fermionic Kitaev wire.

Our analyses are developed within FSS frameworks, which generally
simplify the study of the universal features of critical behaviors,
with respect to studies in the thermodynamic limit.  In the FSS limit
the general requirement of a large length scale $\xi$ of the critical
correlations is not subject to further conditions on the system size
$L$. It only requires that $\xi\sim L$, while critical behaviors in
the thermodynamic limit requires $\xi\ll L$. Therefore much larger
systems are necessary to probe analogous length scales $\xi$ in the
thermodynamic limit.  Equilibrium and out-of-equilibrium FSS behaviors
are often observed for systems of moderately large size, see
e.g. Refs.~\cite{PRV-18,RV-20,RV-21,TV-22,SRDHZ-22}.  Thus FSS
behaviors should be more easily accessed by numerical computations and
experiments where the quantum dynamics can be monitored for a limited
number of particles or spins, such as experiments with quantum
simulators in laboratories, e.g., by means of trapped
ions~\cite{Islam-etal-11, Debnath-etal-16}, ultracold
atoms~\cite{Simon-etal-11, Labuhn-etal-16}, or superconducting
qubits~\cite{Salathe-etal-15, Cervera-18}.

The paper is organized as follows.  In Sec.~\ref{modbath} we present
the fermionic Kitaev wire, equivalent to the quantum Ising chain, and
the model of thermal bath that we use as theoretical laboratory for
our study; we also outline the QQ protocols that we consider and
define the observables to monitor the quantum evolution after
quenching. In Sec.~\ref{scabeh} we outline the out-of-equilibrium
scaling scenarios that are expected to be developed under the dynamic
QQ protocols considered, and support them by numerical computations
for the fermionic Kitaev wires in contact with the thermalizing
bath. Finally, in Sec.~\ref{conclu} we summarize, draw our
conclusions, and add some remarks on the extension of this study to
the dynamic Kibble-Zurek protocols slowly crossing quantum critical
regimes. The appendix reports some details on the numerical
computations for the QQ protocols within fermionic Kitaev wires in
contact with a thermal bath.

\section{Kitaev fermionic wires and thermal baths}
\label{modbath}

\subsection{The fermionic Kitaev chain}
\label{kitaevmod}

We consider fermionic Kitaev wires of $L$ sites with open boundary
conditions, whose quantum unitary dynamics is driven by the
Hamiltonian~\cite{Kitaev-01}
\begin{equation}
  \hat H_{\rm K} = - J \sum_{x=1}^{L-1} \big( \hat c_x^\dagger \hat
  c_{x+1}^{\phantom\dagger} +
  \hat c_x^\dagger \hat c_{x+1}^\dagger+{\rm h.c.}
  \big) - \mu \sum_{x=1}^L \hat n_x ,
  \label{kitaev2}
\end{equation}
where $\hat c_x$ is the fermionic annihilation operator associated
with the site $x$ of the chain, $\hat n_x\equiv \hat
c_x^\dagger \hat c_x^{\phantom\dagger}$ is the particle density
operator.  In the following we assume $J$ as the energy scale, thus we
set $J=1$.

The Hamiltonian~\eqref{kitaev2} can be mapped into a quantum Ising
chain, by means of the Jordan-Wigner transformation, see, e.g.,
Ref.~\onlinecite{Sachdev-book}.  The corresponding spin model is the
quantum Ising chain with open boundary conditions, i.e.
\begin{equation}
  \hat H_{\rm Is} = -\sum_{x=1}^{L-1} \hat \sigma^{(1)}_x \hat
  \sigma^{(1)}_{x+1} - g\, \sum_{x=1}^L \hat \sigma^{(3)}_x,
  \label{isham}
\end{equation}
$\hat \sigma^{(k)}_x$ being the Pauli matrices and $g=-\mu/2$.  In the
following we prefer to stick with the Kitaev quantum wire, because the
thermal baths and observables that we consider are best defined within
the fermionic model. However, the general scaling scenarios that will
emerge apply to both models.

The Kitaev model undergoes a CQT at $\mu=\mu_c = -2$ (corresponding to
$g=g_c=1$ in the quantum Ising chain), between a disordered quantum
phase for $\mu<\mu_c$ (corresponding to $g>1$) and an ordered quantum
phase for $|\mu|<|\mu_c|$ (corresponding to $|g|<1$). Thus, we define
\begin{equation}
  w = \mu - \mu_c = \mu + 2,
  \label{wdef}
\end{equation}
so that one can easily see the correspondence between the Kitaev
Hamiltonian (\ref{kitaev2}) and the generic one reported in
Eq.~(\ref{qudef}), i.e. $\hat{H}_c$ corresponds to the Hamiltonian
(\ref{kitaev2}) for $\mu=\mu_c$, and $\hat{H}_p=-\sum_{x=1}^L \hat
n_x$.  The continuous transition at $w=w_c$ belongs to the
two-dimensional Ising universality class~\cite{Sachdev-book,RV-21},
characterized by the length-scale critical exponent $\nu=1$, related
to the RG dimension $y_w = 1/\nu=1$ of the Hamiltonian parameter
$w$. This implies that, approaching the critical point, the length
scale $\xi$ of the critical quantum fluctuations diverges as $\xi \sim
|w|^{-\nu}$. The dynamic exponent $z=1$ associated with the unitary
quantum dynamics can be obtained from the power law
$\Delta\sim\xi^{-z}$ of the vanishing gap with increasing $\xi$.
Moreover, the RG dimension of the fermionic operators $\hat c_j$ and
$\hat c^\dagger_j$ at the CQT is $y_c = 1/2$, and that of the particle
density operator $\hat n_x$ is $y_n = 1$~\cite{Sachdev-book,RV-21}.

\subsection{Modelization of the thermal bath}
\label{thebath}

In our study we consider a modelization of interaction with a thermal
bath within the Lindblad master equation (\ref{Lindblad}), whose
asymptotic large-time behavior leads to a Gibbs density matrix at a
given finite temperature $T$. In particular, we consider the proposal
developed in Ref.~\cite{DR-21} which applies to quantum models
described by quadratic Hamiltonians, such as that of the fermionic
Kitaev wires. This provides a relatively simple modelization of a
thermal bath leading to thermalization in the large-time limit of the
corresponding Lindblad master equation for the density matrix of the
system.

The Kitaev Hamiltonian (\ref{kitaev2}) with open boundary conditions
can be diagonalized in the Nambu field space by a Bogoliubov
transformation, see e.g. Refs.~\cite{Pfeuty-70,BR-book,DR-21}, so that
we can rewrite it as
\begin{equation}
\hat{H}_{\rm K}=\sum_{k=1}^L\,\omega_k \,\hat b^\dagger_k\, \hat b_k,
  \label{Hdiag}
\end{equation}
where $\omega_k$ are values of the spectrum of the Bogoliubov
eigenoperators $\hat b_k$ (we are neglecting an irrelevant constant
term). Note that both $\omega_k$ and $\hat b_k$ depend on the
Hamiltonian parameter $\mu$.  The relation between the fermionic
operators $\hat{c}_x$ and the Bogoliubov eigenoperators $\hat{b}_k$
can be generally written as~\cite{Pfeuty-70,BR-book,DR-21}
\begin{equation}
  \label{transBogol}
  \hat c_x = \sum_{k=1}^L A_{xk} \,\hat{b}_k + B_{xk}
  \,\hat{b}_k^\dagger,
\end{equation}
where $A$ and $B$ are appropriate $L\times L$ matrices depending on
$\mu$.  Following Refs.~\cite{DR-21,PCR-22}, we write the dissipator
$\mathbb{D}_T[\rho]$ in the Lindblad master equation (\ref{Lindblad})
in terms of the Bogoliubov eigenoperators as
\begin{eqnarray}
  \mathbb{D}_T[\rho] &=&
\sum_k [1-f(\omega_k,T)]
\left( 2 \,\hat{b}_k\,\rho\,\hat{b}_k^\dagger - 
\{\hat{b}_k^\dagger\hat{b}_k,\rho\}\right)  \nonumber\\
&+&
\sum_k f(\omega_k,T)
\left( 2 \,\hat{b}_k^\dagger\,\rho\,\hat{b}_k - 
\{\hat{b}_k\hat{b}_k^\dagger,\rho\}\right), \label{Dtrho}
\end{eqnarray}
where
\begin{equation}
  f(\omega_k,T) = \left( 1 + e^{\omega_k/T}\right)^{-1}.
  \label{fomt}
\end{equation}
When using this homogeneous dissipator term, the Lindblad master
equation (\ref{Lindblad}) ensures the asymptotic large-time
thermalization~\cite{DR-21}. Therefore,
\begin{eqnarray}
  &&\lim_{t\to\infty} \rho(t)  = \rho_t(w,T), \label{asyrho}\\
&&\rho_t(w,T) = \sum_n e^{-E_n(w)/T}|\Phi_n, w\rangle \langle \Phi_n, w|,
 \qquad \label{termrho}
\end{eqnarray}
where $\rho_t(w,T)$ is the density matrix representing the thermal
state, $E_n(w)$ and $|\Phi_n, w\rangle$ are the eigenvalues and
eigenstates of $\hat{H}(w)$.  The asymptotic approach to the thermal
distribution is controlled by the decay-rate parameter
$\gamma$~\cite{DR-21}. Indeed the Liouvillian gap $\Delta_{\cal L}$
that controls the exponential approach to the asymptotic stationary
state of the Lindblad equation is proportional to the decay rate
$\gamma$, i.e.
\begin{equation}
  \Delta_{\cal L}\sim \gamma.
  \label{deltal}
  \end{equation}

The above modelization of thermal baths provides a useful theoretical
laboratory to investigate issues related to the out-of-equilibrium
dynamics in the presence of thermal baths. Its derivation has
been thoroughly discussed in Ref.~\cite{DR-21}. We also mention that it
has been employed in Refs.~\cite{PCR-22,BD-23}.  Some details of the
computations using the Lindblad master equation (\ref{Lindblad}) with
the dissipator (\ref{Dtrho}) are reported in the appendix.

\subsection{Quantum-quench protocols}
\label{proto}

As already anticipated in Sec.~\ref{intro}, we consider two protocols,
differing for the absence or presence of the contact with the thermal
bath during the quantum evolution after quenching, giving respectively
rise to unitary or dissipative dynamics after quenching. We call them
{\em unitary} and {\em dissipative} QQ protocols, respectively.

\begin{itemize}

\item {\em Unitary} QQ protocol: In this simplest QQ protocol the role
  of the thermal bath is limited to that of preparing the initial
  Gibbs state $\rho_t(w_i,T)$ at $t=0$, reported in
  Eq.~(\ref{rhoi}). This can be obtained by keeping the thermal bath
  in contact with the system for a sufficiently long time $t_{\rm th}$, i.e
  $t_{\rm th}\gg \gamma^{-1}$. Then at $t=0$ the Hamiltonian parameter is
  instantaneously quenched from $w_i<0$ to $w\ge 0$ and the thermal
  bath is removed, so that the subsequent time evolution is that of a
  closed fermionic wire, i.e.  it is unitary and only driven by the
  Hamiltonian of the system, cf. Eq.~(\ref{firstprot}).

\item {\em Dissipative} QQ protocol: The quantum evolution starts from
  the same initial Gibbs state $\rho_t(w_i,T)$, but the thermal bath
  is maintained in contact with the system after the QQ from $w_i<0$
  to $w\ge 0$, at $t=0$.  Therefore, the quantum evolution for $t>0$
  is driven by the Lindblad master equation (\ref{Lindblad}) with the
  dissipator term (\ref{Dtrho}). Note that this dynamic protocol
  entails a further time scale $\tau = \gamma^{-1}$, characterizing
  the asymptotic exponential approach to the large-time stationary
  Gibbs state associated with the Hamiltonian $\hat{H}(w)$ and
  temperature $T$.

\end{itemize}

\subsection{Observables monitoring the time evolution}
\label{obs}

To characterize the dynamic properties of the quantum evolution after
the QQ at $t=0$, we consider the subtracted particle-density average
\begin{eqnarray}
n_s(t,L) = {1\over L} {\rm Tr} \left[ \rho(t) \sum_{x=1}^L \hat{n}_x \right]
- n_c(L),
 \label{ddef}
\end{eqnarray}
 where $n_c(L)$ is the ground-state energy density of the Kitaev wire
 of size $L$ at the critical point $w_c=0$ (in the infinite-size limit
 $n_c= 1/2 - 1/\pi$~\cite{Pfeuty-70}). Note that the particle density
 operator $\hat{n}_x$ and the transverse spin component
 $\hat\sigma_x^{(3)}$ of the quantum Ising chain (\ref{isham}) are
 trivially related, indeed $\hat{\sigma}_x^{(3)} = 2 \hat{n}_x$.  In
 the definition of $n_s$, the subtraction of $n_c(L)$ simplifies the
 scaling behavior of $n_s(t,L)$ within the critical regime, cancelling
 the leading analytical behavior~\cite{CPV-14,RV-21}. To monitor the
 spatial correlations, we also consider
\begin{eqnarray}
P(x,y,t) & \!\! = \!\! & {\rm Tr}[\rho(t)\,(\hat c_x^\dagger 
\hat c_{y}^\dagger +
    \hat c_{y} \hat c_{x})],\label{ptf}\\ 
C(x,y,t) & \!\! = \!\! & {\rm Tr}[\rho(t)\, (\hat c_x^\dagger \hat c_{y} 
+ \hat
    c_{y}^\dagger \hat c_{x})].\label{gtf} 
\end{eqnarray}

Some details on the computation of the above quantities during the
time evolution of the QQ protocols are reported in the appendix.

\section{Out-of-equilibrium scaling}
\label{scabeh}

We now discuss the out-of-equilibrium behaviors arising from the QQ
protocols outlined in Sec.~\ref{proto}.  We show that they develop
OFSS behaviors where the effects of the thermal baths are taken into
account by appropriate extensions of the out-of-equilibrium
zero-temperature scaling laws describing soft QQs in
closed systems within their critical regime, already put foward by
earlier works~\cite{PRV-18,RV-21}.

\begin{figure}[!htb]
  \includegraphics[width=0.95\columnwidth]{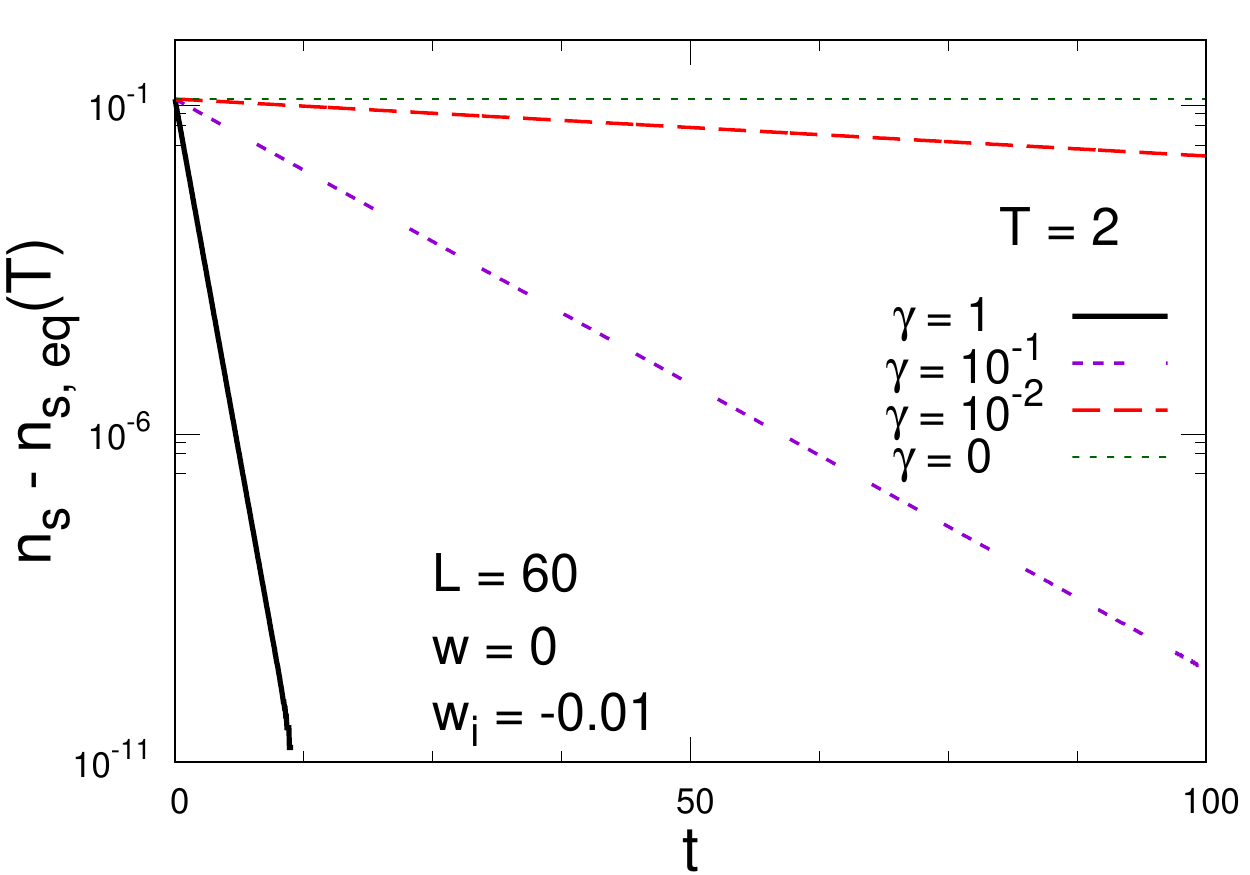}
  \caption{The quantum evolution of the subtracted particle density
    $n_s(t)$, cf. Eq.~(\ref{ddef}), for the dissipative QQ protocol
    entailing a dissipative dynamics after the QQ at $t=0$ of the
    Hamiltonian parameter $w$, describing the persistent interaction
    with the thermal bath, cf.  Eqs.~(\ref{Lindblad}) and
    (\ref{Dtrho}).  These curves refer to a system of size $L=60$,
    temperature $T=2$ of the thermal bath, quenching from $w_i=-0.01$
    to $w=0$, and various values of the decay rate $\gamma$ (the case
    $\gamma=0$ corresponds to the evolution of the close system).  We
    plot the difference $n_s(t,L,T) - n_{s,{\rm eq}}(L,T)$ which is
    expected to vanish in the large-time limit.  In this figure and in
    the following ones, the unity that we use are such that
    $\hslash=1$, $k_B=1$, and $J=1$.}
  \label{plainns}
\end{figure}

The OFSS behaviors that we put forward for QQ protocols considered are
verified by numerical computations for the fermionic Kitaev wire up to
relatively large sizes. See the appendix for details on such
calculations.

As a preliminary example of out-of-equilibriun QQ
behaviors that we want to address, in Fig.~\ref{plainns} we show some
results for the quantum evolution of the subtracted particle density
(\ref{ddef}) along the dissipative protocol outlined in
Sec.~\ref{proto}, after quenching a fermionic Kitaev wire of size
$L=60$, from $w_i=-0.01$ to $w=0$, in the presence of a thermal bath
at a temperature $T=2$, and various values of the decay rate $\gamma$.
The quantum evolution turns out to have a significant dependence on
the decay-rate parameter $\gamma$ that characterized the interactions
between the system and the thermal bath. Indeed, the curves of the
substracted particle density appear to approach its equilibrium value
$n_{s,{\rm eq}}(w=0,T=2)\approx 0.0004601...$ (while at $t=0$ we have
$n_{s,{\rm eq}}(w=w_i,T=2) \approx 0.126598...$), faster and faster
with increasing $\gamma$, actually exponentially as $\exp(-t/\tau)$
with $\tau\sim\gamma^{-1}$, conferming the role of decay rate of the
parameter $\gamma$ within the Lindblad master equation,
cf. Eq.~(\ref{deltal}). Analogous results are obtained for other
observables, such as fermionic correlation functions defined in
Sec.~\ref{obs}. In the following we put forward an out-of-equilibrium
scaling theory for these out-of-equilibrium phenomena within the
quantum critical regime.

\subsection{Zero-temperature scaling in quantum quenches}
\label{zeroT}

We now provide a brief summary of the out-of-equilibrium scaling
theory for close systems, describing QQ protocols within the critical
regime~\cite{PRV-18,RV-21}. The initial state is the ground state
associated with an initial value $w_i<0$, and, after the instantaneous
quench at $t=0$ from $w_i$ to $w$, the quantum evolution is driven by
the Schr\"odinger equation.

Out-of-equilibrium scaling laws can be obtained by extending those
valid at equilibrium, allowing for a time dependence essentially
controlled by the time scaling variable $\Theta \sim t\,\Delta$, which
is obtained by assuming that the relevant time scale of the critical
modes is proportional to the inverse energy difference $\Delta$ of the
lowest states. We refer to Ref.~\cite{RV-21} for a through
presentation of the scaling arguments leading to the asymptotic OFSS
behaviors.

Let us consider the out-of-equilibrium evolution (after quenching) of
generic observables, such as the expectation value $O$ at time $t$ of
a local operator $\hat{O}({\bm x})$ and its fixed-time correlations
$G_O=\langle \hat{O}({\bm x}) \hat{O}({\bm y})\rangle$. The general
working hypothesis underlying out-of-equilibrium FSS frameworks is
that the expectation value of $\hat{O}({\bm x})$ and its correlation
functions obey asymptotic homogeneous scaling laws~\cite{RV-21}, such
as
\begin{eqnarray}
O(t, {\bm x}, L, w_i, w) \approx  b^{-y_o} {\cal O}(t/b^z,
{\bm x}/b, L/b, b^{y_w} w_i, b^{y_w} w),\quad
    \label{Oscaquefss0}
\end{eqnarray}
  where $b$ is an arbitrary (large) length scale, $y_o$ is the RG
  dimension of the local operator $\hat{O}_{\bm x}$ and the RG
  exponents $y_w$ and $z$ are determined by the universality class of
  the CQT (they are the RG dimensions of the Hamiltonian parameter $w$
  and the temperature $T$, respectively). Thus both the initial and
  final values of $w$, i.e.  $w_i$ and $w$, take the same RG exponent
  $y_w$, being coupled to the RG perturbation ${\hat H}_p$ within the
  Hamiltonian. Note that we do not assume translation invariance,
  which is generally broken by the presence of boundaries, such as
  those arising from open boundary conditions.

OFSS can be straightforwardly derived by fixing $b=L$ in the above
homogenous scaling law. Then, we expect the OFSS of the expectation
value $O$ of a generic local operator $\hat{O}_{\bm x}$, of its
spatial average $\hat{O}_a =L^{-d}\sum_{\bm x}\hat{O}_{\bm x}$, and
its two-point correlation function $G_O$, develop the asymptotic OFSS
behavior~\cite{PRV-18,RV-21}
\begin{eqnarray}
    && O(t, {\bm x}, L, w_i, w) \, \approx \, L^{-y_o} \,
  {\cal O}(\Theta, {\bm X}, \Phi_i, \Phi),
\nonumber\\
  && O_a(t, L, w_i, w) \, \approx \, L^{-y_o} \, {\cal O}_a(\Theta,
\Phi_i, \Phi),   \label{Oscaquefss}\\
&& G_{O}(t, {\bm x}_1, {\bm x}_2, L,
  w_i, w) \, \approx \, L^{-2y_o} \, {\cal G}_{O}(\Theta, {\bm X}_1,
  {\bm X}_2, \Phi_i, \Phi), \nonumber
\end{eqnarray}
where the scaling variables appearing in the scaling functions ${\cal
  O}$, ${\cal O}_a$, and ${\cal G}_O$ are defined as
\begin{equation}
  \Theta\equiv\frac{t}{L^z},\;\; {\bm X}_i \equiv \frac{{\bm x}_i}{L},\;\;
  \Phi_{i} \equiv
L^{y_w} \, w_i , \;\; \Phi \equiv L^{y_w} \,w.
  \label{scalvarque}
\end{equation}
The OFSS limit is obtained in the large-$L$ and large-$t$ limit
keeping the above scaling variables fixed. These conditions ensure
that the system remains within the universal critical regime during
the quantum evolution.  Note that in the scaling law
(\ref{scalvarque}) the dynamic features are essentially encoded in the
time dependence of the scaling variable $\Theta\sim t\,\Delta$.  The
other features, in particular when $w_i=w$, are analogous to those
arising from equilibrium FSS at CQTs~\cite{CPV-14,RV-21}, where the
argument $\Phi=L^{y_w} w$ of the scaling functions is controlled by
the RG dimension $y_w$ of the relevant parameters $w$ at the RG fixed
point associated with the CQT.

The above OFSS equations can be straightforwardly applied to the
observables defined in Sec.~\ref{obs}, after a quench from $w_i$ to
$w$ at $t=0$, keeping into account that the RG dimension of the
subtracted particle density is $y_n = 1$, and that of the fermionic
operator $\hat c_x$ is $y_c=1/2$.  Note that the dominant analytical
contributions to the particle density~\cite{CPV-14,RV-21} coming from
the analytical background are canceled in the difference $n_s$ defined
in Eq.~(\ref{ddef}), whose leading asymptotic behavior arises from the
quantum critical modes, therefore it is analogous to that of $O_a$ in
Eq.~(\ref{Oscaquefss}), with $y_o=y_n$.  Analogously one can apply the
OFSS in Eq.~(\ref{Oscaquefss}) to observables and correlation
functions constructed with the spin operators of the quantum spin
chain (\ref{isham}).  The OFSS functions are expected to be universal
with respect to the microscopic details of the model, apart from
nonuniversal multiplicative rescaling and normalizations of its
arguments.  Within isolated fermionc Kitaev wires and quantum Ising
chains, the OFSS arising from soft QQs has been verified by numerical
computations for various boundary conditions, and also along their
quantum first-order transition line~\cite{PRV-18,RV-21}.

The OFSS limit is expected to be approached with power-law suppressed
corrections.  There are various sources of scaling corrections when
approaching the OFSS. Of course, they include those that are already
present at equilibrium. In particular, the irrelevant RG perturbations
are sources of scaling corrections for the asymptotic behavior of the
free-energy density~\cite{PV-02,RV-21}.  In the case of
one-dimensional quantum systems undergoing CQTs belonging to the
two-dimensional Ising universality class, the leading scaling
corrections from irrelevant RG perturbations are suppressed as
$L^{-\omega}$ with $\omega=2$~\cite{CHPV-02,CPV-14}. However, other
contributions may become more relevant~\cite{PV-02,CPV-14,RV-21}, such
as those arising from the presence of analytical backgrounds, from the
presence of boundaries (which generally gives rise to $O(1/L)$
corrections), and, in the case of correlation functions, from RG
mixings of the source fields [this for example happens in the case of
  the correlation functions of the fermionic field $\hat{c}_x$, for
  which corrections are $O(1/L)$].  These scaling corrections have
been confirmed by numerical results~\cite{CPV-14,RV-21}. Therefore, we
expect that the asymptotic OFSS of fermionic Kitaev wires and quantum
Ising chains with open boundary conditions is generally approached
with $O(1/L)$ corrections.

\subsection{OFSS along the unitary QQ protocol}
\label{scalprota}

For the simplest unitary protocol reported in Sec.~\ref{proto}, where
the quantum evolution is that of the isolated fermionic wire, the
request that the dynamics remains within the critical regime implies
that the temperature of the initial Gibbs state must be appropriately
suppressed in the large-$L$ OFSS limit, to obtain a nontrivial
out-of-equilibrium critical limit.  This is analogous to what happens
within the equilibrium FSS, where one introduces the scaling
variable~\cite{SGCS-97,Sachdev-book,RV-21}
\begin{equation}
  \Xi \equiv L^z T,
 \label{Xidef}
\end{equation}
to allow for a nonzero temperature in the FSS of the observables.
Therefore, like equilibrium FSS, we conjecture that the temperature of
the initial Gibbs state enters the OFSS associated with the unitary QQ
protocol by adding a further dependence on $\Xi$ in the scaling
functions (\ref{Oscaquefss}).  In other words, a nontrivial asymptotic
OFSS limit is expected to be realized in the large-$L$ and large-$t$
limits keeping also $\Xi$ fixed, beside the scaling variables already
defined in Eq.~(\ref{scalvarque}).  Therefore, we expect that the OFSS
of standard QQ protocols starting from ground states,
cf. Eq.~(\ref{Oscaquefss}), changes into
\begin{eqnarray}
   O(t, {\bm x}, L, w_i, w, T) \, \approx \, L^{-y_o} \, {\cal
    O}(\Theta, {\bm X}, \Phi_i, \Phi, \Xi),
    \label{Oscaquefssa}
\end{eqnarray}
and analogously for its spatial average $O_a$ and the correlation
function $G_O$.

\begin{figure}[!htb]
  \includegraphics[width=0.95\columnwidth]{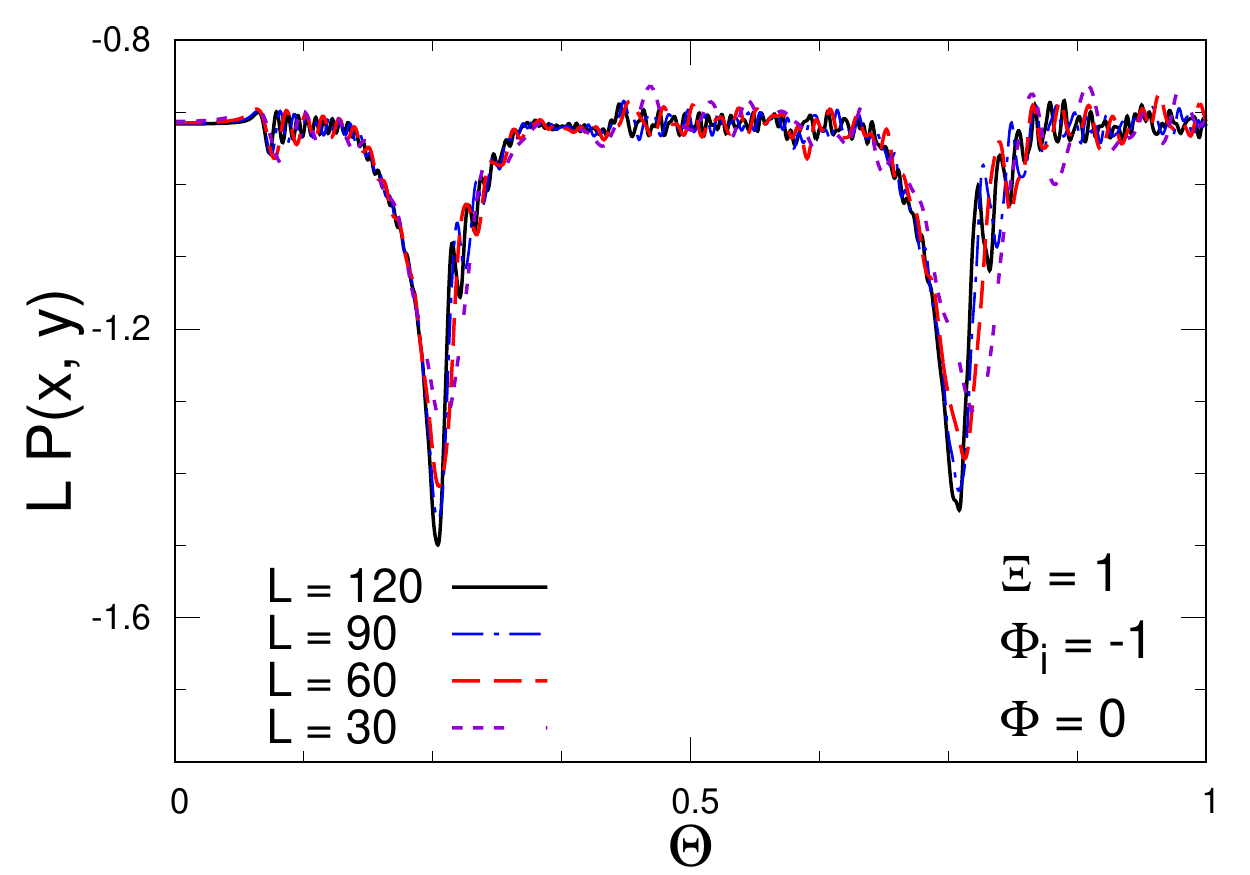}
  \includegraphics[width=0.95\columnwidth]{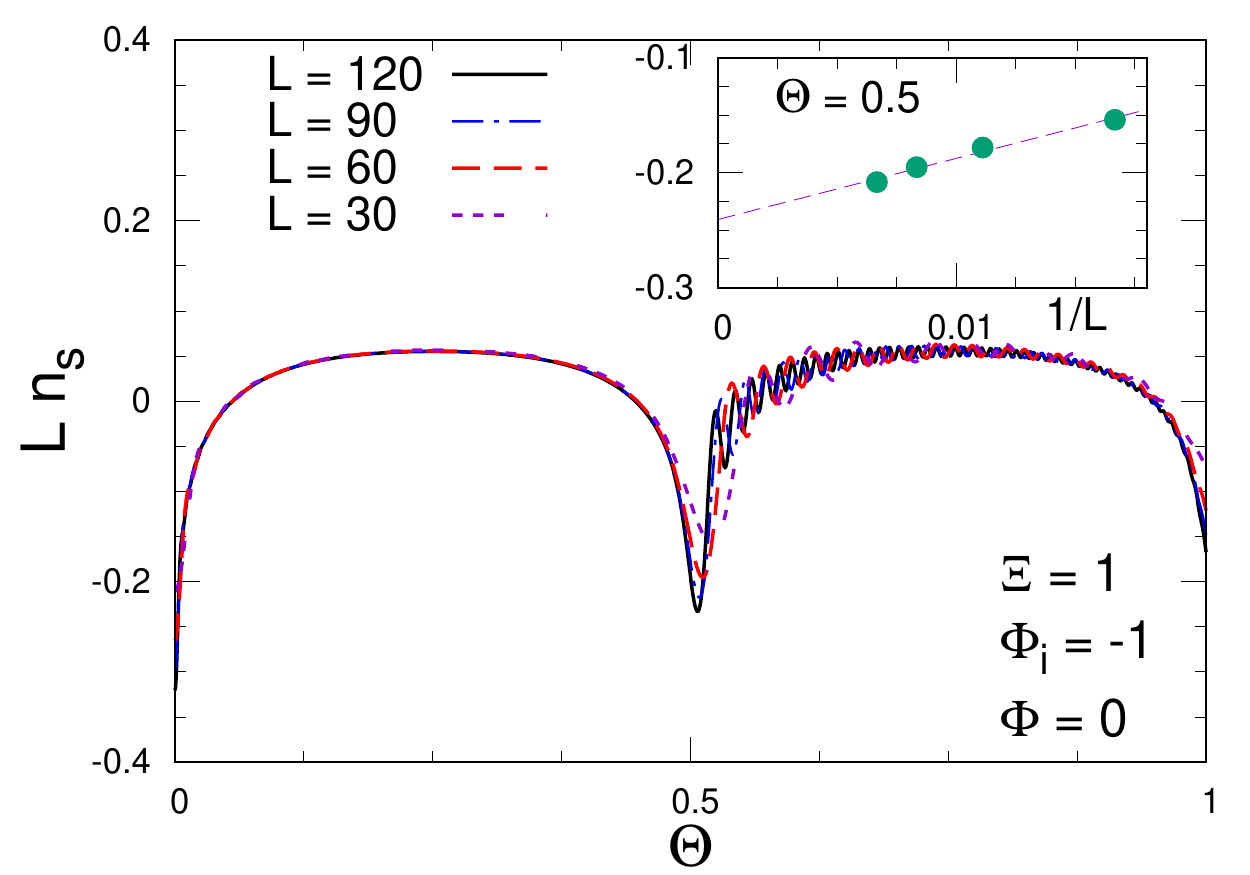}
  \caption{OFSS behavior of the subtracted particle density (bottom)
    and the fermionic correlation function $P(x=L/3,y=2L/3,t)$,
    cf. Eq.~(\ref{ptf}), arising from the unitary QQ protocol, for
    various lattice sizes $L$, at fixed $\Xi=L^z T =1$,
    $\Phi_i=L^{y_w} w_i=-1$ and $\Phi=L^{y_w} w=0$, versus the time
    scaling variable $\Theta=t/L^z$.  These computations nicely
    support the OFSS behaviors reported in
    Eq.~(\ref{Oscaquefssa}). The inset of the bottom figure shows that
    the approach to the OFSS limit is consistent with $O(1/L)$
    corrections.  Analogous results are obtained for other values of
    the scaling variables.}
  \label{protares}
\end{figure}

The numerical analysis for the fermionic Kitaev wire under the unitary
protocol fully support to this OFSS, obtained by extending the QQ FSS
behaviors of closed systems starting from an initial ground
state. This is clearly demonstrated by the curves reported in
Fig.~\ref{protares}, associated with the quantum evolutions of the
subtracted particle density $n_s(t)$ and the fermionic correlation
$P(x,y,t)$ (the other fermionic correlation $C(x,y,t)$ develops an
analogous OFSS).

\subsection{OFSS along the dissipative QQ protocol}
\label{scalprotb}

We now discuss the dynamics arising from the dissipative protocol
outlined in Sec.~\ref{proto}, when the quantum evolution after
quenching is described by the Lindblad master equation
(\ref{Lindblad}) with the thermal-like dissipator (\ref{Dtrho}), to
modelize the interaction with a thermal bath characterized by a
temperature $T$ (which does not change after quenching)
and decay rate $\gamma$.

\begin{figure}[!htb]
  \includegraphics[width=0.95\columnwidth]{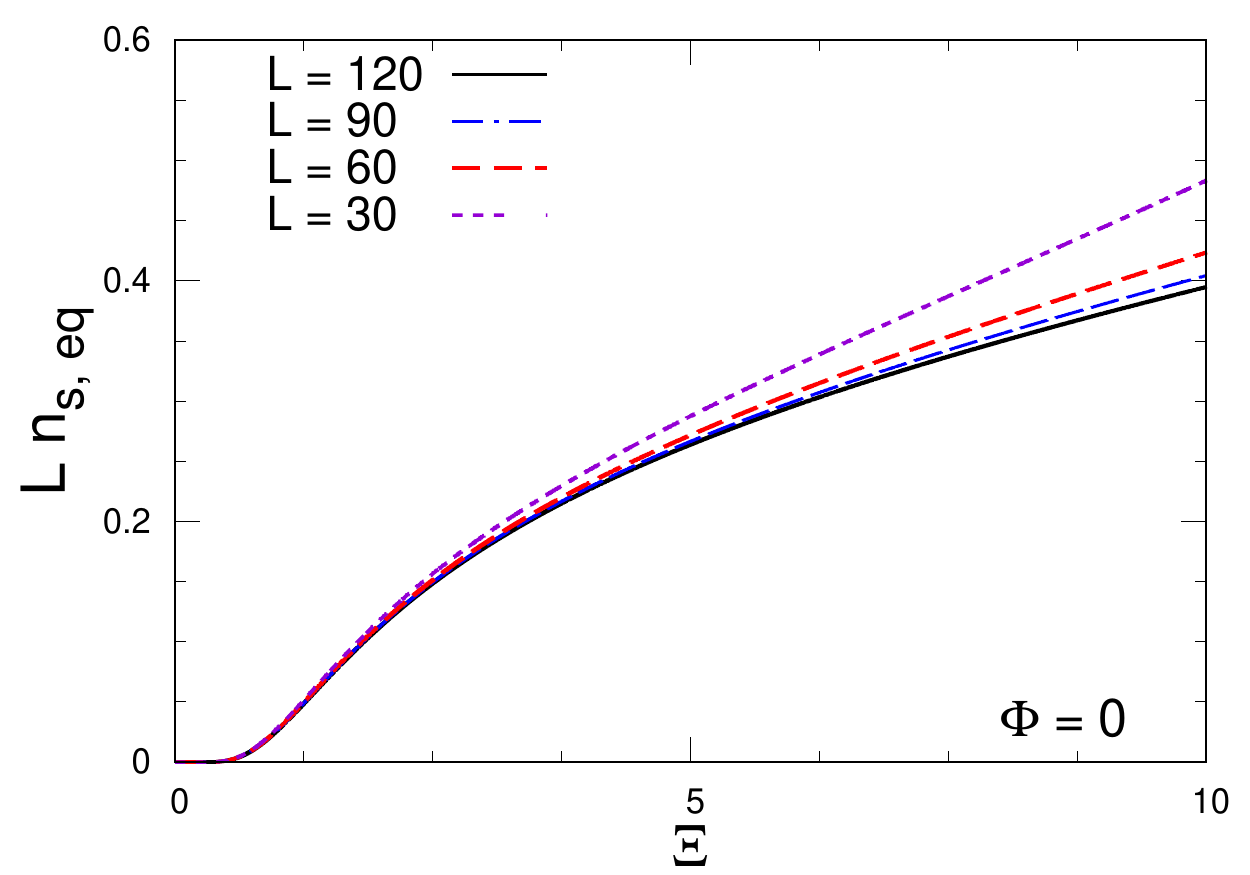}
  \caption{ Equilibrium FSS of the subtracted particle density
    $n_{s,{\rm eq}}$ at the critical point $w=0$, versus the rescaled
    temperature $\Xi=L^z T$. With increasing $L$, the data show the
    expected convergence to the equilibrium FSS reported in
    Eq.~(\ref{nseqsca}) with $y_n=1$. }
  \label{eqns}
\end{figure}

\begin{figure}[!htb]
  \includegraphics[width=0.95\columnwidth]{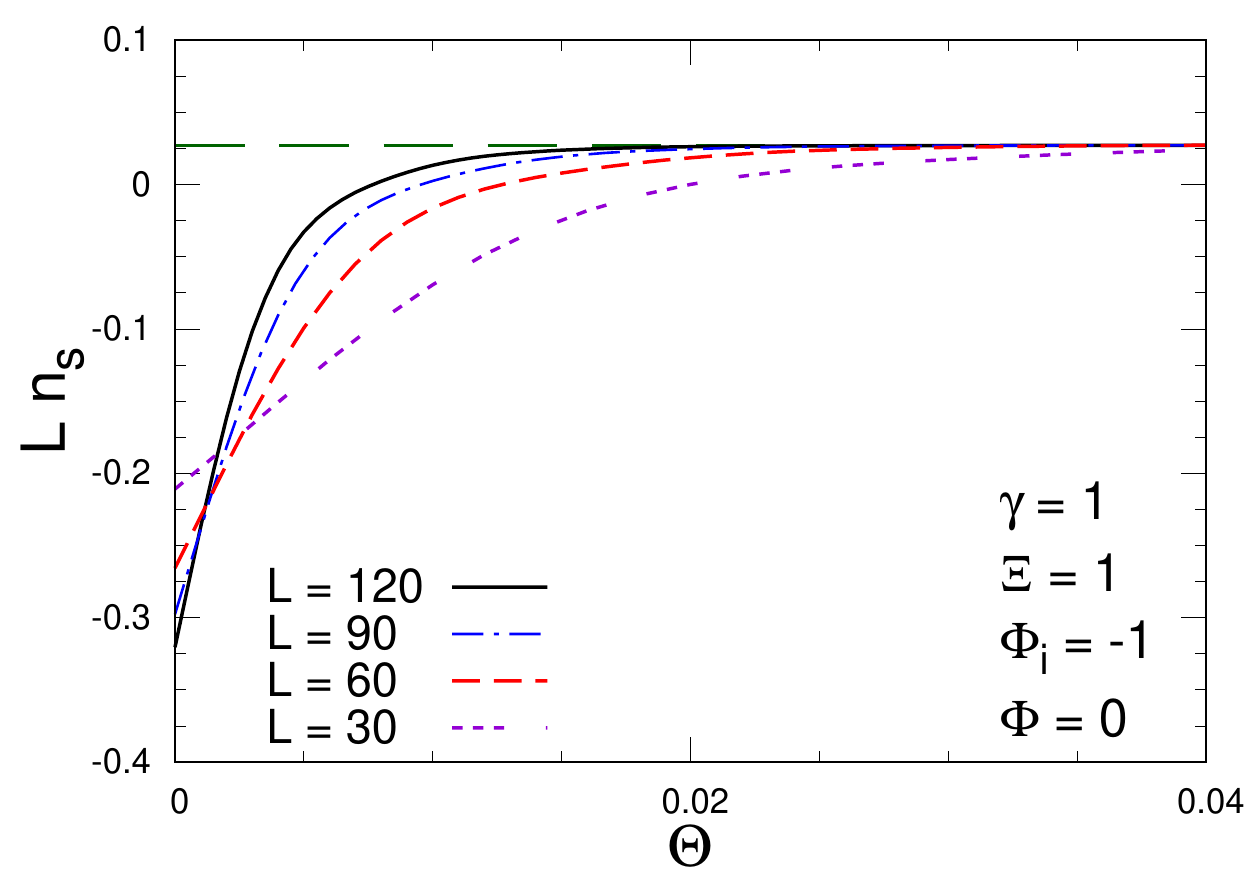}
  \caption{Quantum evolution of the subtracted particle density
    arising from the dissipative QQ protocol, when rescaling all
    quantities involved in the quench protocol, except for the decay
    rate $\gamma$. With increasing $L$, the curves appear to approach
    the equilibrium FSS value at finite temperature (where the
    temperature dependence enters through the scaling variable
    $\Xi=L^z T$) faster and faster, reflecting a nonuniform
    convergence for any $\Theta>0$. The dashed line shows the
    equilibrium value of $n_s$ for $\Phi=0$ and $\Xi=1$, which is
    asymptotically approached by the various curves.  }
  \label{protbresgamma}
\end{figure}

We expect that the temperature $T$ of the thermal bath must be
rescaled as in the case of the unitary QQ protocol, i.e. we must
consider again the associated scaling variable $\Xi$ already defined
in Eq.~(\ref{Xidef}).  However, since the QQ moves the system
out-of-equilibrium, also the decay rate $\gamma$, and corresponding
time scale $\tau=\gamma^{-1}$, associated with the interactions with
the thermal bath is expected to play a relevant role to establish a
corresponding nontrivial OFSS limit.  This was already noted in
Ref.~\cite{BD-23} in the analysis of dynamic protocols entailing the
variation of the temperature at the critical point.

When keeping $\tau$ constant in the FSS limit where the
scaling variable $\Theta=t/L^z$ is kept fixed, in the large-$L$ limit
we have eventually that
\begin{equation}
  t = \Theta \, L^z \gg \tau,
  \label{tggtau}
  \end{equation}
which is the condition ensuring thermalization for any finite value
$\Theta>0$. Therefore, when keeping $\tau$ fixed, the quantum
evolution is not expected to develop a nontrivial OFSS limit. Indeed,
in the large-$L$ limit, the system turns out to suddenly approach an
equilibrium Gibbs state (associated with the Hamiltonian parameter $w$
and temperature $T$) with respect to the rescaled time $\Theta$,
without any further relevant evolution of the system for any
$\Theta>0$.  Therefore, if the temperature is rescaled by keeping
$\Xi=L^z T$ fixed, we must recover the equilibrium FSS behavior in the
presence of a thermal bath at temperature $T$, such as that associated
with the subtracted particle density~\cite{CPV-14,RV-21}
\begin{equation}
  n_{s,{\rm eq}}(w,L,T) \approx L^{-y_n} {\cal N}(\Phi,\Xi),
  \label{nseqsca}
  \end{equation}
where $\Phi=L^{y_w} w$, and the temperature dependence enters through
the associated scaling variable $\Xi=L^z T$.  In Fig.~\ref{eqns} we
show some equilibrium data at the critical point $w=\Phi=0$, versus
$\Xi$, showing the approach to the asymptotic large-$L$ equilibrium
FSS (\ref{nseqsca}).  The realization of the equilibrium FSS within
the QQ protocol at fixed $\gamma$ is demonstrated by the plots
reported in Fig.~\ref{protbresgamma}, which show the somewhat trivial
convergence toward the equilibrium FSS for any finite $\Theta>0$.

The above results suggest that also the the decay rate $\gamma$ of the
system-bath interactions must be rescaled to observe a nontrivial OFSS
limit as a function of the time scaling variable $\Theta$, to create
the conditions for a balanced competition between the critical
Hamiltonian driving and the interactions with the thermal bath. As
already put forward in the case of other homogeneous dissipative terms
in the Lindblad equation~\cite{NRV-19-cd,RV-20-kz,TV-21,RV-21,FT-23},
for example associated with particle-decay or particle-pumping
dissipative mechanisms, a nontrivial OFSS limit is obtained by
rescaling the decay rate of the dissipative term, so that the scaling
variable
\begin{equation}
  \Gamma \equiv L^z \gamma \sim \gamma/\Delta
  \label{gammadef}
\end{equation}
is kept fixed in the OFSS limit, where $\Delta$ is the energy
difference of the lowest eigenstates of $\hat{H}(w)$ at the critical
point $w=w_c=0$.  Then an OFSS behavior emerges from the nontrivial
competition between the critical unitary dynamics and the dissipative
driving arising from the thermal bath.

In conclusion, on the basis of the above scaling arguments, the OFSS
arising from the dissipative QQ protocols in the presence of a thermal
bath is expected to be given by
\begin{eqnarray}
   O_a(t, L, w_i, w, T, \gamma) \, \approx \, L^{-y_o} \,
  {\cal O}_a(\Theta, \Phi_i, \Phi, \Xi, \Gamma),\quad
  \label{Oscaquefssprotb1}
\end{eqnarray}
and
\begin{eqnarray}
&& G_{O}(t, {\bm x}_1, {\bm x}_2, L, w_i, w, T, \gamma)  \, \approx
    \label{Oscaquefssprotb2} \\
    &&\qquad\qquad
     L^{-2y_o} \, {\cal G}_{O}(\Theta, {\bm X}_1, {\bm X}_2, \Phi_i,
    \Phi, \Xi, \Gamma). \qquad \nonumber
\end{eqnarray}
In the large-$\Gamma$ limit the above OFFS behaviors at fixed $\Xi$ is
expected to approach the corresponding equilibrium FSS, faster and
faster in terms of $\Theta$, matching the behavior at finite $\gamma$.
Moreover, we also expect that the equilibrium FSS is also approached
in the large-$\Theta$ limit at fixed $\Gamma$ and $\Xi$, independently
of $\Gamma$, but faster and faster with increasing~$\Gamma$.

\begin{figure}[!htb]
  \includegraphics[width=0.95\columnwidth]{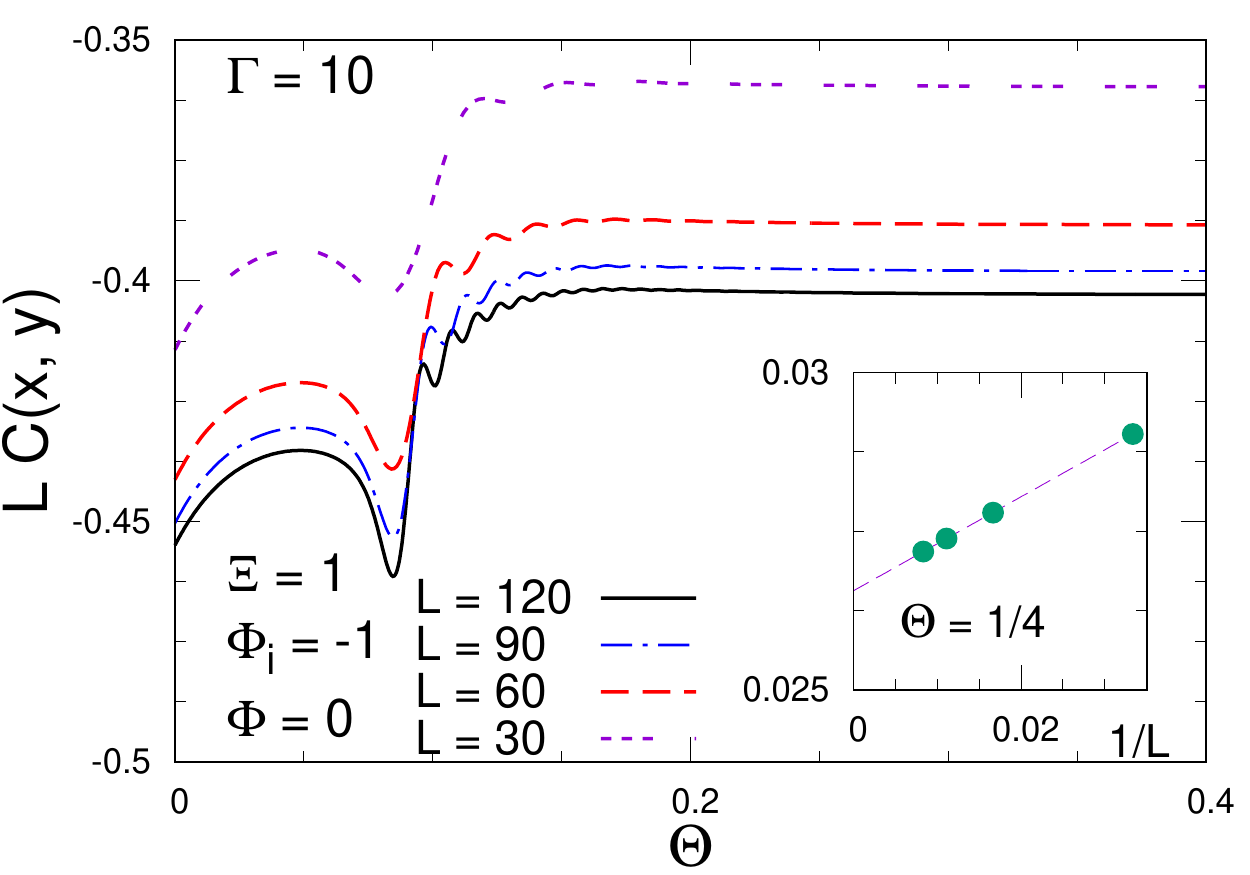}
  \includegraphics[width=0.95\columnwidth]{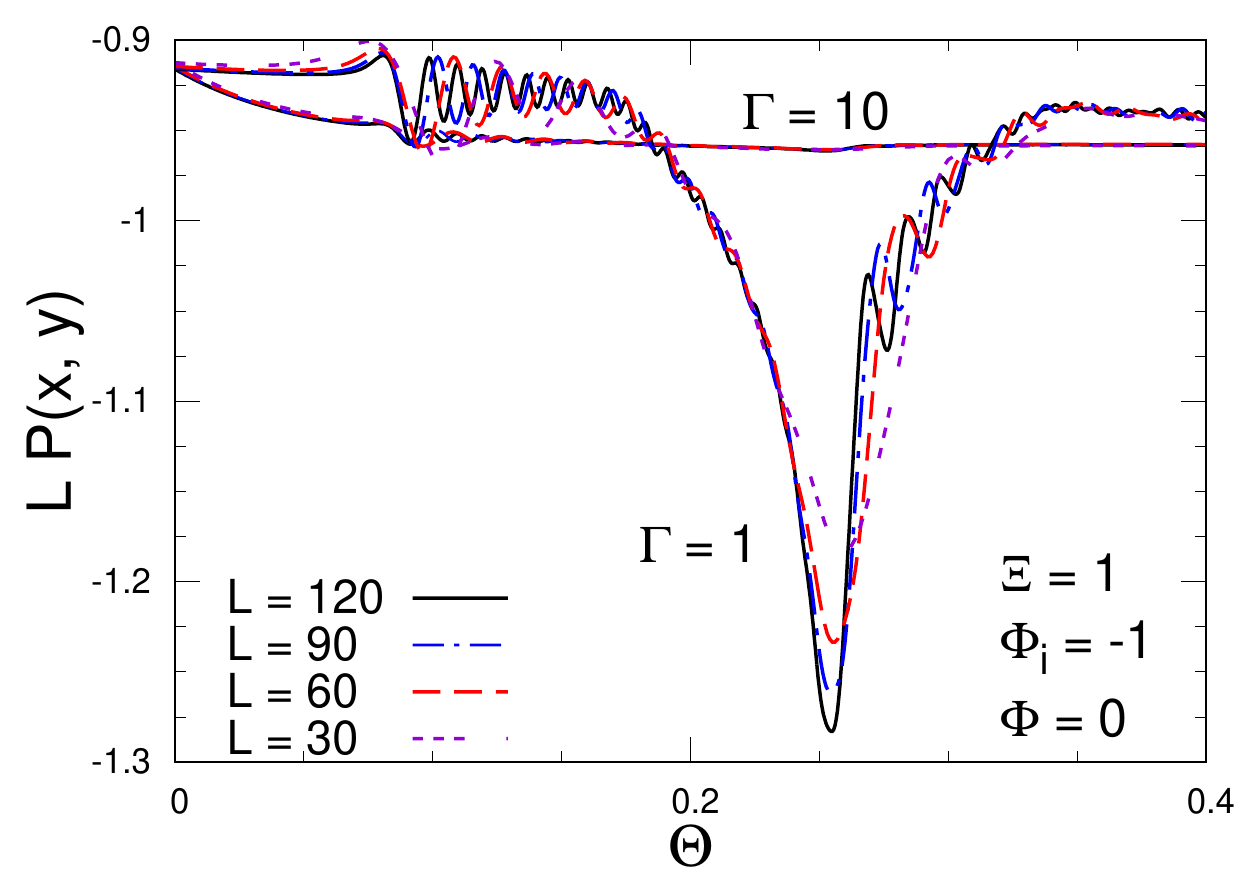}
    \includegraphics[width=0.95\columnwidth]{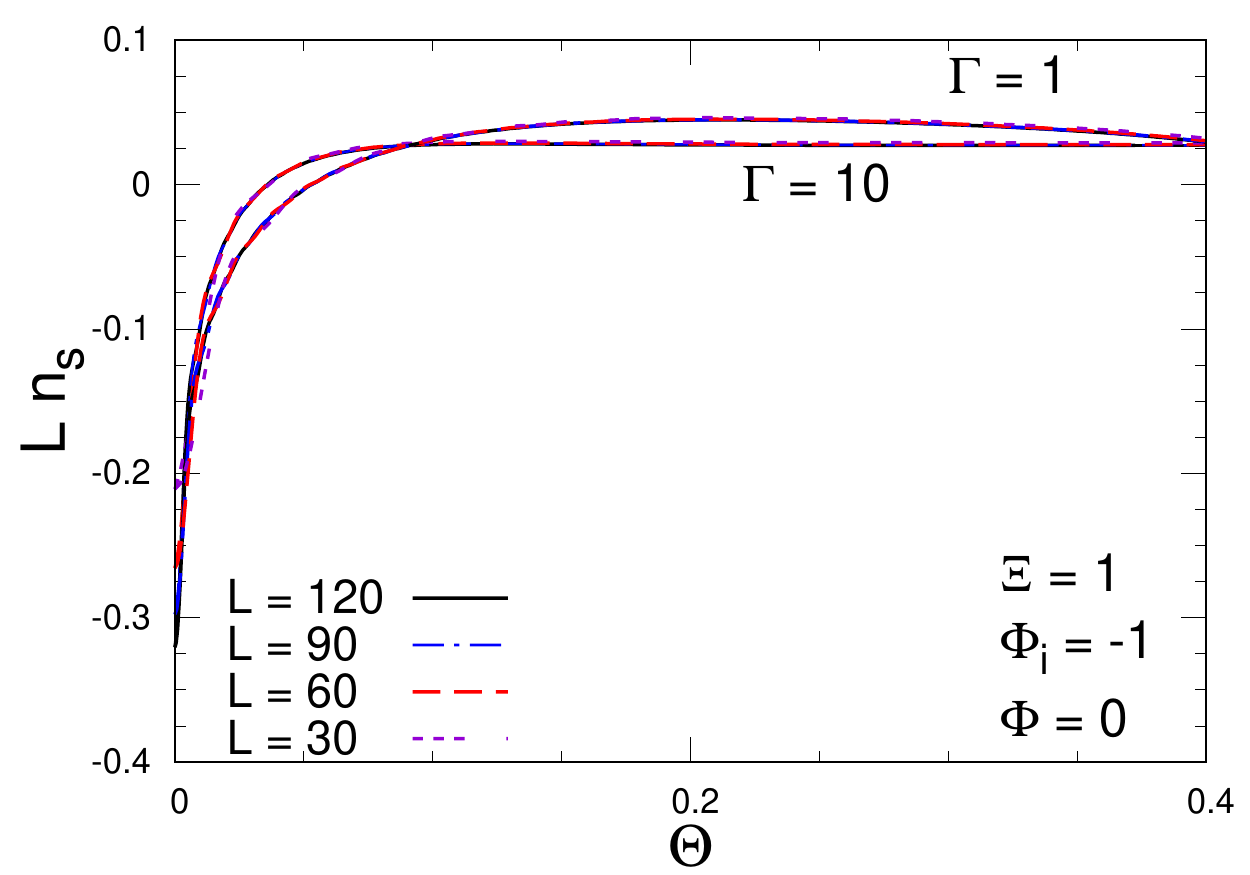}
    \caption{Quantum evolutions along the dissipative protocol, fully
      supporting the OFSS reported in Eqs.~(\ref{Oscaquefssprotb1})
      and (\ref{Oscaquefssprotb2}). We report curves for $L\,n_s$
      (bottom), $L\,P(x=L/3,y=2L/3,t)$ (middle), and
      $C(x=L/3,y=2L/3,t)$ (top), for various values of $L$, at fixed
      $\Phi_i=-1$, $\Phi=0$, $\Xi=1$, and two values of
      $\Gamma=L^z\gamma$, i.e.  $\Gamma=1,\,10$ (except for the top
      figure where we only report data for $\Gamma=10$ to ensure a
      good readability).  The inset of the top figure shows that the
      OFSS is approached with $O(1/L)$ corrections. Analogous results
      are obtained for other values of the scaling variables. }
  \label{protbresgammaresc}
\end{figure}

Again, the numerical results for the particle density $n_s(t)$ and
correlation functions $P$ and $C$ fully support the above OFSS
equations, i.e. Eq.~(\ref{Oscaquefssprotb1}) for $n_s(t)$ with
$y_o=y_n=1$, and Eq.~(\ref{Oscaquefssprotb2}) for $P$ and $C$ with
$y_o=y_c=1/2$. Some results are reported in
Fig.~\ref{protbresgammaresc}.  We also stress that analogous results
are expected for other observables, for example the correlation
functions of the spin operator of the equivalent formulation provided
by the quantum Ising chains.

\section{Conclusions}
\label{conclu}

We have reported a study of the effects of thermal baths to the
out-of-equilibrium dynamics of many-body systems within their quantum
critical regime close to a zero-temperature CQT. In particular, we
analyze the out-of-equilibrium quantum evolution arising from QQs of
the Hamiltonian parameters within two different protocols involving a
thermal bath coupled homogeneously to the system.  Within the first
protocol, named unitary QQ protocol, the thermal bath is used to
prepare the system at $t=0$ in a finite-temperature Gibbs state, then
the dynamics after quenching of the Hamiltonian parameters is assumed
unitary, i.e., the thermal bath is removed during the quantum
evolution for $t>0$.  The second protocol, named dissipative QQ
protocol, starts from the same initial condition, but the thermal bath
is not removed after quenching, and the quantum evolution for $t>0$ is
assumed to be described by the Lindblad master equation
(\ref{Lindblad}).  The dissipative term of the Lindblad equation is
supposed to simulate a thermal bath, such that the many-body system is
driven to a large-time finite-temperature Gibbs state.  This
dissipative protocol is characterized by a further time scale
$\tau=\gamma^{-1}$, related to the decay rate of the interactions
between the system and the bath.

Within OFSS frameworks, we argue that, when the thermal baths are
associated with a sufficiently small temperature, their effects can be
taken into account by appropriate extensions of the zero-temperature
out-of-equilibrium scaling laws describing soft QQs of isolated
systems within the critical regime.  For the unitary QQ protocol,
where the thermal bath only determines the initial Gibbs state and the
evolution is unitary, a nontrivial OFFS limit is simply obtained by
rescaling the temperature as $T\sim L^{-z}$, similarly to equilibrium
FSS.  Along the dissipative QQ protocol, where the thermal bath is not
removed after quenching, the dynamics is more complicated, and the
decay rate $\gamma$ plays a relevant role.  Indeed, in addition to the
rescaling of the temperature $T$ associated with thermal bath, one
also needs to rescale $\gamma$ as $\gamma\sim L^{-z}$ to obtain a
nontrivial OFSS. Otherwise, when keeping $\gamma$ fixed, the dynamics
converges toward the equilibrium FSS at finite temperature, which
happens suddenly after quenching with respect to the time scale
$t_c\sim L^z$ of the critical regime.  Therefore the scaling behavior
when keeping $\gamma$ fixed becomes somehow trivial, reproducing the
equilibrium FSS for any rescaled time $\Theta = L^{-z} t >0$ in the
large-$L$ limit.

Our scaling arguments are supported by numerical results with the
paradigmatic fermionic Kitaev model, or equivalently quantum Ising
chain, at its CQT separating quantum disordered and ordered phases.
We consider a particular modelization of the thermal bath that
guarantees the asymptotic thermalization within the Lindblad
formulation of the dynamics of open systems.  However, we note that
the scaling arguments used to arrive at the OFSS laws for critical QQs
are general, and therefore we expect that the emerging
out-of-equilibrium scenarios also apply to many-body systems at generic
CQTs in contact with homogenous thermal baths, in any dimension.

We finally remark that the out-of-equilibrium scaling arguments we put
forward, leading to the OFSS of QQs in the presence of a thermal bath,
can be extended to other protocols giving rise to out-of-equilibrium
dynamics. Another interesting class of dynamic protocols entails slow
variations of the Hamiltonian parameters across the critical regime of
a quantum transition, such as those associated with the quantum
Kibble-Zurek (KZ) problem~(see
e.g. Refs.~\cite{Kibble-76,Kibble-80,Zurek-85,Zurek-96,
  ZDZ-05,PG-08,Dziarmaga-10,PSSV-11,CEGS-12, Dutta-etal-book,
  RDZ-19,RV-21,DV-23}).  In standard KZ protocols starting from the
ground state for an initial parameter $w_i<0$, the out-of-equilibrium
quantum evolution arises from the linear time dependence of one
Hamiltonian parameter, $w(t) = t/t_s$ in Eq.~(\ref{qudef}), where
$t_s$ is the time scale of the KZ protocol. Since $w(t)$ crosses the
critical point at $t=0$, the system passes through the quantum
critical regime, moving it away from equilibrium even in the
large-$t_s$ limit, and developing a peculiar out-of-equilibrium
scaling behaviors.  In particular, the interplay between the size $L$
of the system and the time scale $t_s$ of the protocol develops OFSS
behaviors~\cite{RV-21,DV-23} when $t_s\to \infty$ and $L\to\infty$,
keeping the scaling variables $\Omega_t \equiv
t/t_s^{\kappa}=t/t_s^{z/(y_w+z)}$ and $\Upsilon\equiv t_s/L^{y_w+z}$
(thus $\Omega_t=t/t_s^{1/2}$ and $\Upsilon=t_s/L^2$ for the fermionic
Kitaev wire or quantum Ising chain) fixed.

KZ-like protocols can be also extended to systems interacting with a
thermal bath, such as that outlined in Sec.~\ref{thebath}, starting
from a Gibbs state for an initial $w_i<0$ and the temperature $T$ of
the thermal bath.  Then we may consider a time evolution driven by the
Lindblad master equation (\ref{Lindblad}), with a time-dependent
Hamiltonian $\hat H[w(t)]$ and the dissipator term (\ref{Dtrho}),
where also the Bogoliubov operators are assumed to be time dependent
to adapt themselves to the time dependence of $w$.  Analogously to the
OFSS of QQs in contact with thermal baths, to define a nontrivial OFSS
limit in KZ protocols, we expect that both the temperature $T$ and the
decay rate $\gamma$ associated with the bath must be rescaled, as
$T\sim L^{-z}$ and $\gamma\sim L^{-z}$.  If only the temperature of
the thermal bath is rescaled as $T \sim L^{-z}$, while $\gamma>0$ is
kept fixed, the time interval associated with a variation of
$\Omega_t$ in the KZ scaling limit, i.e. $\Delta_\Omega t \sim
t_s^{\kappa}\Delta\Omega_t$, becomes eventually much larger than the
time scale $\tau\sim\gamma^{-1}$ of the interaction with the thermal
bath.  Since $\tau/\Delta_\Omega t\to 0$ in the KZ limit, the system
effectively thermalizes at each rescaled time $\Omega_t$.  Therefore,
in the KZ limit the quantum evolution is expected to pass through
equilibrium finite-temperature states, thus effectively resulting into
adiabatic evolutions reproducing the equilibrium finite-temperature
FSS as a function of $L^{y_w}w(t)$.  Therefore, like dissipative QQ
protocols, the observation of a nontrivial OFSS in KZ protocols
requires the simultaneous rescaling of the time scale $\tau$
associated with the interaction with the thermal bath.  The necessary
rescaling of the decay rate $\gamma$ of the dissipative term in the
Lindblad master equation has been also put forward for KZ protocols in
the presence of other dissipative mechanisms, such as those related to
particle decay or pumping~\cite{RV-20-kz}.

\acknowledgments

We thank Giulia Piccitto and Davide Rossini for interesting and useful
discussions.

\appendix

\section{Details on the computations}
\label{detcomp}

In this section we provide some details of the computations for the
fermionic Kitaev wire in the presence of a thermal bath.

\subsection{Asymptotic thermal states}
\label{asyther}

The dynamics of the system in contact with the thermal bath described
by the Lindblad master equation (\ref{Lindblad}) with the dissipator
term (\ref{Dtrho}) leads to thermal states, such as those described by
  the density matrix reported in Eq.~(\ref{termrho}).  To compute the
  correlation functions of the fermionic operators $\hat{c}_x$ in
  thermal states of the Hamiltonian $\hat{H}(w)$, one can use the
  relation with the Bogoliubov eigenoperators $\hat{b}_k$,
  cf. Eq.~(\ref{transBogol}), and the thermal correlations of the
  Bogoliubov operators $\hat{b}_k$, i.e.
 \begin{eqnarray}
      \braket{b^\dagger_k b_q} \equiv {\rm Tr}[\rho_t(w,T) b_k^\dagger
        b_q] = \frac{\delta_{kq}}{1 + e^{\omega_k/T}},
\label{bcoth}
    \end{eqnarray}
    corresponding to the standard Fermi-Dirac distribution function.
    Note also that the other correlations $\braket{b_k b_q}$ and
    $\braket{b_k^\dagger b_q^\dagger}$ vanish.  Then the correlation
    functions of the original fermionic field $\hat{c}_x$ can be
    straightforwadly obtained from Eq.~(\ref{transBogol}).

\subsection{Computations for the unitary protocol}
\label{uniprotcomp}

In the unitary QQ protocol, one starts from a Gibbs state associated
with the Hamiltonian parameter $w_i$ and the temperature $T$, then at
$t=0$ one instantaneously changes $w_i\to w$ and removes the contact
with the thermal bath. Therefore the quantum evolution is unitary,
described by the Schr\"odinger equation (\ref{firstprot}). One may
easily obtain closed equations for the evolution of the correlation
functions $C$ and $P$ defined in Eqs.~(\ref{ptf}) and (\ref{gtf}).

We introduce the correlations
\begin{eqnarray}
  \label{RedCorr}
  \mathscr{C}_{x,y} = {\rm Tr}\Bigr[\rho(t) \hat c^\dagger_x
    \hat c_y\Bigr],\quad
  \mathscr{P}_{x,y} = {\rm Tr}\Bigr[\rho(t) \hat c^\dagger_x
    \hat c^\dagger_y\Bigr],
\end{eqnarray}
whose quantum evolution can be written as
\begin{align}
\frac{d\mathscr{C}_{x,y}}{dt}& =
i\,\bigr[\mathscr{C}_{x,y+1} - \mathscr{C}_{x-1,y} +
\mathscr{C}_{x,y-1} - \mathscr{C}_{x+1,y} \bigr] - \notag\\
-i\, \Bigl(& \mathscr{P}_{y,x-1}^\dagger -
\mathscr{P}_{y,x+1}^\dagger \Bigl)  + i\, \Bigl(
\mathscr{P}_{x,y-1} - \mathscr{P}_{x,y+1} \Bigl), \\
\frac{d\mathscr{P}_{x,y}}{dt} &=
-i\,\bigr[\mathscr{P}_{x,y+1} + \mathscr{P}_{x+1,y}+
\mathscr{P}_{x,y-1} + \mathscr{P}_{x-1,y} \bigr] - \notag\\
&- 2\,i\,\mu  \,\mathscr{P}_{x,y}-i\,\Bigl(
\delta _{x-1,\,y} - \delta _{x+1,\,y} \Bigl) - \notag \\
 &- i\, \Bigl( \mathscr{C}_{x,y-1} -
\mathscr{C}_{y,x-1} - \mathscr{C}_{x,y+1}
+ \mathscr{C}_{y,x+1} \Bigl).
\end{align}
The initial conditions are easily obtained by the relations with the
thermal correlations of the Bogoliubov operators associated with the
initial Gibbs state. Then the fermionic correlation function are
obtained by
\begin{eqnarray}
C(x,y,t) = 2\,{\rm Re}{\mathscr{C}_{x,y}}(t),\;\;
P(x,y,t) = 2\,{\rm Re}{\mathscr{P}_{x,y}}(t).\quad
\label{cpcpmp}
\end{eqnarray}
The above differential equations are solved using the four-order
Runge-Kutta method. The particle density is obtained from the data of
$\mathscr{C}_{x,x} = {\rm Tr}\Bigr[\rho(t) \hat c^\dagger_x \hat
  c_x\Bigr]$.

\subsection{Computations for the dissipative protocol}
\label{dissprotcomp}

For the dissipative QQ protocol, where the thermal bath is kept in
contact with the system, the evolution is driven by the Lindblad
master equation (\ref{Lindblad}), which can be equivalently written in
terms of the time dependence of Heisenberg operators $\hat{O}_{\rm H}(t)$,
i.e.~\cite{PCR-22, DR-21}:
\begin{eqnarray}
   \label{EQLindblad}
   \partial_t \hat{O}_{\rm H}(t)   = i\,\Bigr[\hat{H}(w),
     \hat{O}_{\rm H}(t) \Bigr] + \gamma
   \widehat{\mathbb{D}}_T[\hat{O}_{\rm H}(t)],
   \label{oheq}
\end{eqnarray}
   where 
\begin{eqnarray}
&& \widehat{\mathbb{D}}_T[\hat{O}_{\rm H}(t)] = \sum _k f(\omega_k)\,\biggr[2\hat
    b_{k}^\dagger \hat{O}_{\rm H}(t) \hat b_{k}- \Bigl\{ \hat{O}_{\rm
      H}(t), \, \hat b_{k} \hat b_{k}^\dagger \Bigl\} \biggr]
  \nonumber \\ && + \sum_k (1-f(\omega_k)) \,\biggr[ 2\hat b_{k}
    \hat{O}_{\rm H}(t) \hat b_{k}^\dagger - \Bigl\{ \hat{O}_{\rm
      H}(t), \, \hat b_{k}^\dagger \hat b_{k} \Bigl\}
    \biggr],\qquad\qquad
\label{dohdef}
\end{eqnarray}
where $\hat b_k$ are the Bogoliubov operators associated with the
Hamiltonian $\hat{H}(w)$.

The initial state at $t=0$ is the Gibbs state for the Hamiltonian
parameter $w_i$. This state corresponds to the steady state solution
of the Eq.~(\ref{EQLindblad}) with $\hat{H}(w_i)$. Then, the change of
the Hamiltonian parameter to $w\neq w_i$ leads to a change of the
Bogoliubov operators diagonalizing the Hamiltonian.  We call
$\{b'_k\}$ the operators which diagonalizes $\hat{H}(w)$,
\begin{equation}
  \label{quenchHdiag}
  \hat{H}(w)=\sum _{k=1}^L\,\omega'_k \,\hat b'^\dagger _k\,
  \hat b'_k,
\end{equation}
where $\{\omega'_k\}$ is the Bogoliubov spectrum associated with
$\hat{H}(w)$.
To evaluate the correlations of the Bogoliubov operatore $\{b'_k\}$,
one can solve the Eq.~(\ref{EQLindblad}) for couples of operators
$\{b'_k\}$, obtaining~\cite{DR-21}
\begin{eqnarray}
  && \braket{b'^\dagger_k b'_k} = ( 1 - e^{-2 \gamma t})
  f(\omega'_k) +
  e^{-2\gamma t} \braket{b'^\dagger_k b'_k}_0 ,\nonumber \\
 && \braket{b'^\dagger_k b'_q} =
e^{i(\omega'_k - \omega'_q)t-2 \gamma t}
  \braket{b'^\dagger_k b'_q}_0,
  \nonumber\\
&&  \braket{b'^\dagger_k b'^\dagger_q} =
e^{i(\omega'_k + \omega'_q)t-2 \gamma t}
  \braket{b'^\dagger_k b'^\dagger_q}_0,
  \nonumber \\
  &&  \braket{b'_k b'_q} =
  e^{-i(\omega'_k + \omega'_q)t-2 \gamma t}
  \braket{b'_k b'_q}_0.
  \label{EQLindbladprime}
\end{eqnarray}

The initial values $\braket{b'^\dagger_k b'_q}_0$ of the correlations
is computed on the initial Gibbs state associated with $w_i$, and it
can be obtained using the relations between $\{b_k \}$ to $\{b'_k\}$.
This relation can be formally derived as follows~\cite{DR-21}.
Introducing the fermionic Nambu field $\mathbb{C}^\dagger =
(\hat{c}_1^\dagger, ...,\hat{c}_L^\dagger, \hat{c}_1,...,\hat{c}_L)$,
their relations with the Bogoliubov operators $\mathbb{B}(w)^\dagger =
(\hat{b}_1^\dagger, ...,\hat{b}_L^\dagger, \hat{b}_1,...,\hat{b}_L)$
corresponding to the Hamiltonian $\hat{H}_{\rm K}(w)$ are obtained by
a unitary transformation, $\mathbb{C} = \mathbb{T}(w)\mathbb{B}(w)$.
See e.g. Ref.~\cite{DR-21} for more details.  Therefore one can formally
derive the relation between the Bogoliubov operators $\hat{b}_k'$ and
$\hat{b}_k$, corresponding to the Hamiltonian parameters $w_i$ and $w$
respectively, from the general relation
\begin{equation}
\mathbb{B}(w_2)= \mathbb{T}(w_2)^\dagger \mathbb{T}(w_1) \mathbb{B}(w_1).
  \label{hatbrel}
  \end{equation}

Finally, to compute the time-dependent observables defined in
Sec.~\ref{obs}, one can use the relations between the fermionic
correlation functions associated with $\hat{c}_x$ and those of the
Bogoliubov operators $\hat{b}_k$, such as
  \begin{eqnarray}
   C(x,y) &=& \sum_{k,q=1}^L \Big[
  A^*_{xk}A_{yq} \braket{b^\dagger_k b_q}
  + B^*_{xk} B_{yq} \braket{b_k b^\dagger_q} \nonumber \\
  &&+ A^*_{xk} B_{yq} \braket{b^\dagger_k b^\dagger_q} +
  B^*_{xk} A_{yq} \braket{b_k b_q}\Big]
    \label{initcorr}
\end{eqnarray}
where $A$ and $B$ are the matrices entering Eq.~(\ref{transBogol}).

\end{document}